\documentclass{article}
\usepackage{arxiv}

\usepackage[utf8]{inputenc} 
\usepackage[T1]{fontenc}    
\usepackage{hyperref}       
\usepackage{url}            
\usepackage{booktabs}       
\usepackage{amsfonts}       
\usepackage{nicefrac}       
\usepackage{microtype}      
\usepackage{lipsum}
\usepackage{cite}
\usepackage{amsmath,amssymb,amsfonts}
\usepackage{algorithmic}
\usepackage{graphicx}
\usepackage{textcomp}
\usepackage{todonotes}
\usepackage{xcolor}
\usepackage{graphicx}
\graphicspath{ {./images/} }

\title{A Distributed `Black Box' Audit Trail Design Specification for Connected and Automated Vehicle Data and Software Assurance}

\author{
 Gregory Falco$^1$ \\
  Freeman Spogli Institute\\
  Stanford University\\
  Stanford, CA 94305\\
  \texttt{falco@stanford.edu} \\
   \And
 Joshua E. Siegel$^1$ \\
  Department of Computer Science and Engineering\\
  Michigan State University\\
  East Lansing, MI 48824 \\
  \texttt{jsiegel@msu.edu} \\
}

\begin{document}
\maketitle
\footnotetext[1]{Both authors contributed equally to this manuscript.}

\begin{abstract}
Automotive software is increasingly complex and critical to safe vehicle operation, and related embedded systems must remain up-to-date to ensure long-term system performance. Update mechanisms and data modification tools introduce opportunities for malicious actors to compromise these cyber-physical systems, and for trusted actors to mistakenly install incompatible software versions. A distributed and stratified ``black box'' audit trail for automotive software and data provenance is proposed to assure users, service providers, and original equipment manufacturers (OEMs) of vehicular software integrity and reliability. The proposed black box architecture is both layered and diffuse, employing distributed hash tables (DHT), a parity system and a public blockchain to provide high resilience, assurance, scalability, and efficiency for automotive and other high-assurance systems.
\end{abstract}


\section{Introduction}
Vehicle complexity is increasing, with a growing volume of software and number of sensors, actuators, compute modules, and intra- and extra-vehicular networks dedicated to enhancing vehicle performance, safety, reliability, and efficiency\cite{siegel_itsj}.

While automotive \textit{hardware} is typically immutable, recently, \textit{software} upgrades -- whether dealership-installed or over-the-air updates -- have begun to enhance and enable features within an extant fleet. Today, software controls critical vehicle functions including Advanced Driver Assistance Systems (ADAS) and highly-automated driving. These applications are trained and tested for safety on particular vehicles, with software variants designed for specific hardware configurations, operator use cases, or even tailored to individual drivers' behavior. Because this software is both highly varied and safety-critical, there is a need to assure software provenance, integrity, and immutability to ensure that the proper versions are installed. Having the correct software is necessary in order to assure that the overall system performs as-designed.  

Sensors, actuators, and vehicle-specific configuration and persistent state data are similarly sensitive to intentional and incidental data interference. Sensor messages and actuation requests may be spoofed\cite{nie2017free}, causing local system failure -- or, in the case of a connected vehicle, the misbehavior of a larger transportation network. The risk of large-scale failure is significant, with $75$\% of new vehicles expected to be connected by 2020\cite{halder2019secure}. Data shared car-to-car might be used to automate braking, or to reroute vehicles to avoid a traffic jam, and while verified data may stabilize traffic and reduce accidents\cite{ortega2018trusted}, ``bad'' data may lead to the converse. These software and data integrity challenges are complicated by new business models such as shared mobility, wherein vehicles may be exposed to untrustworthy individuals and potentially-skilled adversaries.

As a first step towards addressing these challenges, we propose a method for tracking and validating critical persistent and semi-permanent data (e.g. vehicle configuration parameters, identifiers, and odometery metrics stored within EEPROM) and software (version and variant data and update history) within vehicles. Our solution resembles the audit trail functionality found in a flight data recorder (FDR) or 'black box' \cite{grossi1999aviation}, but distributed and layered in nature. We employ distribution for resilience in the case of an attack on or malfunction of vehicular components, our proposed design is resource-efficient with respect to in-car computing and network-loading constraints\cite{siegel_itsm} to ensure scalability and compatibility with the incumbent vehicle fleet, and it has the potential to later be adapted to assure integrity of sensor readings and actuator commands in addition to providing software version control and configuration assurance. We note that broader shifts in security culture must accompany any technological solution, and that the envisioned technique is not a comprehensive solution for assuring vehicle security -- rather, it is complementary to the larger security-focus gaining traction in industry. 

The proposed distributed black box implements a lightweight Distributed Hash Table (DHT), networked datastores that utilize consensus voting to validate data to log any sensitive data or software change event, including updates to software version, variant, changes to vehicle identification data, or significant events including major increments of vehicle miles travelled. In high-assurance systems, the DHT's content is backed up by a parity system so that critical data will not be lost should an embedded device, inter- or intra-vehicular network, or memory module fail. Multiple DHT nodes regularly feed information into a blockchain that may be simply queried by an OEM without having to ping multiple nodes of the DHT to access routing tables. The proposed solution is hardware- and communications-agnostic and capable of running within the hardware and network constraints imposed by a ``typical'' modern vehicle context, including limitations brought about by the use of cost-efficient embedded systems with constrained computation, memory, and connectivity. 

The findings may be generalized to other embedded and constrained systems, and extended to other types of data (e.g. environmental context data to better understand accidents, sensor or actuator data for keyframe validation, or as part of a proof-of-location system). It may also be implemented at the dealership- and OEM-level to create a record of fleet-wide software installations to support and improve warranty claim handling, service records, and to provide data useful for insurers, law enforcement, fleet managers, and component suppliers. 

Our architecture is proposed as being complementary to contemporary approaches to software and data system assurance, rather than as a fully-functional and all-encompassing solution. Our distributed black box enables an audit trail of software changes and data modification and is one of many best practices (e.g. resilient software design, end-to-end communication encryption, etc.) that collectively enable assurance for complex cyber-physical systems such as connected and automated vehicles. Standards such as ISO 26262 \cite{iso26262} and NIST's Cyber-Physical Systems Framework \cite{cps_framework} offer more comprehensive guidance for improving the security of embedded systems.

In Background, we provide information about the capabilities, constraints, and challenges in assuring data and software integrity in an automotive context. Prior Art details how blockchain has been used to solve vehicular data integrity challenges and describes unmet needs, particularly stemming from the computational intensity of blockchain implementations. We propose design considerations for a resilient audit trail in a like-named section, and then provide an architectural overview of the distributed black box. We describe alternative approaches considered and highlight why our chosen solution best addresses the problem landscape for connected and automated vehicles as well as other high-assurance systems, before closing with newly-enabled capabilities and a future vision in Enabled Capabilities and Future Opportunities.

\section{Background}\label{background}
This section describes automotive computing and network architectures as well as contemporary approaches used in industry related to software updates, version management, and data integrity. This motivates the need to improve data integrity and software version control, as well as identifies available resources and constraints within ``typical'' vehicles.

\subsection{Automotive Network and Computing Architecture}
A modern vehicle might feature hundreds of sensors, dozens of low-to-mid-power embedded compute modules locally controlling various functions, and one or two higher-power computers interconnected by more than ten distinct networks. Often, these networks are bridged by one or more gateway devices, and increasingly, at least one module provides access to the Internet through one or more extra-vehicular networks such as Bluetooth, cellular 4G or 5G, or Dedicated Short-Range Communication (DSRC).\cite{siegel_itsj}

Notably, both the computing nodes and the network capability of modern vehicles is constrained due to power, cost, and other design requirements. The resulting system features low-clocked processors, limited memory, and low bandwidth. Worsening this problem is the long service life of vehicles -- averaging $11.6$ years in the United States, as of 2016\cite{polk_age} -- and the fact that vehicle hardware is rarely upgraded, and rather replaced only upon its failure. 

\subsection{Automotive Software and Data: Authenticity and Version Control}
In vehicles, software is increasingly used to control the performance of critical systems such as acceleration, braking, and steering. This software must perform hard-realtime tasks, degrade gracefully, and safely handle spurious data and service interruptions\cite{iso26262} and it is therefore validated extensively throughout the development process. While software may initially be robust and bug-free, the late addition of vehicle-differentiating features may introduce bugs or create unanticipated and untested use cases. To address these challenges, one or multiple compute modules must be updated with appropriate, authentic software to ensure that the entire vehicle electronic system is validated as being self-compatible.  

It is therefore necessary to track software versions across in-vehicle modules to ensure software compatibility and therefore safety and performance. The problem is not as simple as loading the latest version onto each module with every update -- software must be configured differently based on the vehicle's options, with the resulting variant reflecting the availability of different sensors, actuators, and messages transiting across the various in-vehicle networks. Further complicating factors, even the same option codes may implement components from different production lots or even suppliers and therefore require different calibration or interface code. 

At the time of assembly, these parameters are known and it is straightforward to program the vehicle with a tested and approved software payload. Only certain configurations are allowable from the factory, and modules are programmed at the time of assembly, with the manufacturer recording the (hardware and) software version to a database. Once deployed, and when bugs are identified or features are added, there are three methods for software updates: dealership programming, over-the-air (OTA) updates, and do-it-yourself updates. These modalities are more difficult to secure. 

When a vehicle is brought in for service, critical updates are typically downloaded to in-car modules using a specialized programming tool. The dealership or service center's programming tool often relies upon on-line software capable of querying the manufacturer's database of available software, and that database contains information about software's compatibility and interoperability with other software and hardware within the vehicle. Based on these data, the module's software is updated at the same time as its dependencies, including software installed on other computing devices. The OEM and/or supplier may optionally record this programming event to a database along with the vehicle identification number, providing an auditable history of programming events. 

However, not all service centers use this software, and as a result, manufacturers may not have full insight into the software version installed on individual vehicles. For example, some aftermarket programming tools may only update a single module without regard for vehicle options or software version interoperability. Complicating this further is the advent of Over-The-Air (OTA) updates, which can be used to improve security and to reduce service costs \cite{10.1007/978-3-319-66972-4_12,halder2019secure}, but which may not necessarily check compatibility or provide confirmation of successful software installation. There is also a growing market for aftermarket software\cite{koebler2017american} and hobbyist reprogramming (``car hacking'') that manufacturers are unable to track, as the OEM or supplier's online database is never queried and the vehicle may not necessarily identify that a programming event has taken place, e.g. in the case diagnostic services are used to rewrite specific memory locations, rather than a wholesale code re-flashing event. Similarly, end-users may buy and install pre-programmed modules from scrapped vehicles or part houses assuming they are plug-and-play while in fact the software versions are incompatible, leading to safety concerns.  

Despite these challenges, Original Equipment Manufacturers (OEMs) have an obligation to maintain vehicle safety for the entire product lifecycle\cite{halder2019secure}. Being able to accurately identify installed software and its provenance for all in-vehicle modules could reduce illegitimate warranty claims and limit liability exposure stemming from improper software configuration for manufacturers and their suppliers. Similar techniques could be used to identify whether a ``variant configuration'' (software-enabled features) have been changed by unauthorized users, or whether ``immutable'' parameters, like mileage, airbag status, service event history, or the Vehicle Identification Number, have been changed without authorization.

We aim to solve the problem of automotive software provenance validation and assuring integrity of critical data for the incumbent and future vehicle fleet by creating a black box audit trail capable of monitoring data and software versions that is resistant to contemporary attack modes, including in-situ data modification, aftermarket software updates, and module swapping. If sufficiently lightweight and performant, such a system could be extended to a variety of use cases including validating critical actuator or sensor data or capturing context relating to accidents. 

\section{Prior Art}\label{prior_art}
Researchers have previously called for and discussed the merit of in-vehicle data recorders (IVDR) for highly-automated systems and robotics \cite{perez2010argos, winfield2017case,wortham2017robot, yao2020smart}. Such calls did not however present design specifications for a distributed black box that have the aforementioned capabilities.

For time-of-production version control, OEMs often utilize networked databases to record hardware and software metadata for the initially-installed payload. Similarly, OEMs may run online, Internet-enabled programming services to better-control access to software and to allow for the creation of a database, which maps software versions to vehicle identification numbers and associated configuration data. 

During programming events, the software may modify memory on multiple electronic control units as a form of distributed ledger -- that is, the radio, engine controller, and instrument panel computer may all store common metadata such as software version, vehicle identification, airbag status or crash data, and mileage, to make code modification more difficult by requiring simultaneous manipulation. 

This approach is vulnerable to two common attacks: the first is EEPROM modification, wherein data are modified directly on one or more device's memory with the battery removed. The second is through ``official'' programming mechanisms, such as Universal Diagnostic Services or J2534 reprogramming, but with users directly modifying memory addresses or using offline and/or cracked software to modify otherwise-``protected'' memory. With this type of reprogramming, many modules will not register a software-change event, meaning the installed code and the on-device metadata may reflect different versions, leading to a false sense of security and/or poor interoperation. 

To enhance the protection of module programming and of critical memory, there have been efforts to use blockchain to secure elements of vehicle software, data, and communications. For example, blockchain has been used to secure and validate software updates\cite{steger2018secure,dorri2017blockchain,Chanson:2017:BPE:3123024.3123078,10.1007/978-3-319-66972-4_12}, to secure odometer readings and vehicle identification\cite{brousmiche2018digitizing,chanson2017blockchain}, and for credential management\cite{malik2018blockchain}, data management, and vehicle authentication\cite{moinet2017blockchain}. Blockchain has also been used to store and secure vehicle lifecycle data, service records, and accident histories and reconstruction from supply chain through end-of-life\cite{brousmiche2018digitizing,anwar2019ensuring,8493118,8626103}. In some implementations, vehicles mine blocks themselves, and region-bounded ledgers may be used\cite{SINGH2018219}. In-vehicle blockchains have also been proposed\cite{10.1007/978-3-030-12839-5_47,alam2018securing}. 

In general, these approaches use a heavy-weight blockchain implementation. While this may be suitable for future expensive and purpose-built shared vehicles, which amortize up-front costs across a larger number of rides, such an approach may be incapable of running on the constrained hardware present in the incumbent vehicle fleet. What's needed is the ability to store these data at large scale without requiring significant computational overhead for the embedded devices themselves. We propose one possible architecture suitable for use in validating vehicle software versions and in-car data integrity for certain, safety-critical parameters. 

\section{Design Considerations}\label{design}
Ensuring compatibility across the diverse embedded systems found in contemporary vehicles requires three foundational design requirements for the distributed black box:

First, the embedded devices will need to self-identify a hash representing software and hardware metadata that may be uniquely-identifying. This may include: design date, date and location of manufacture, version number, variant code, serial number, and/or VIN. The hash will be sent to a master computing unit with heightened capabilities, such as the infotainment system head unit. In many vehicles, such a device already exists - often running embedded or full versions of consumer operating systems. The head unit will report the unique hashes externally for analysis. The hash identifier is crucial considering each embedded device's maintenance requirements may be dependent on a combination of factors reflected in the metadata. For example, devices with the same software version may have different update requirements because of variations in so-called ``identical'' devices produced by different suppliers, or the same supplier in different lots. At all times there should be a running log of hash identifiers present in the vehicle that is held within the master unit.

Second, the communication integrity of the embedded system metadata is critical, moreso than the privacy of the data itself. If an attacker is able to change the log maintained of the embedded systems versions, it is possible to introduce software incompatibilities that would result in unsafe operations. This challenge is particularly significant for connected and automated vehicles, as their computational payload is more substantial than conventional vehicles, and their drive-by-wire capabilities create a unique attack vector subject to compromise resulting from data modification or corruption not present in some human-operated vehicles. Even minor data alterations could cause an autonomous vehicle to be taken out of service, though the solution we present is equally applicable to conventional, human-operated vehicles and their computing payloads. Also, providing a robust mechanism for reporting the time and state of software versions during reporting could be leveraged for other use cases requiring traceable, temporal histories, such as proving location.

Finally, the master unit that holds the version control information will need to be able to communicate with other high-compute units and the OEM via wireless telemetry. Such communication could be direct from the master node or indirect after traversing multiple trusted intermediaries over the network to reach the server. Here again, many contemporary vehicles directly link a master unit such as an infotainment head unit with a telemetry module capable of providing high-speed and low-latency extravehicular connectivity. 

One potential system-level implementation is described below.

\section{Architectural Overview}\label{architecture}
We propose a layered distributed architecture that employs both a distributed hash table (DHT) at the in-vehicle embedded system level that feeds into a blockchain light client to communicate externally to achieve the outlined objectives.

A distributed hash table is a distributed ledger that can store and retrieve data efficiently using a key across a network of nodes. No one node contains all of the system's data at a given time, therefore, it can be extremely memory efficient as compared with a blockchain. The embedded systems will host a DHT that can be distributed across the vehicle's internal network (level 1). 

Because each node of the DHT does not hold the complete ledger which will need to be accessed by the OEM, each  DHT node will communicate with a single blockchain light client that resides in the master unit (level 2).

The light client that operates over the Etherum blockchain \cite{reilly2019smart} will serve as a central location within the vehicle for the otherwise distributed embedded system data to be collected. The light client subsequently reports the embedded system data to a designated full blockchain node housed by the vehicle's OEM (level 3).

The light client will communicate this data at pre-defined intervals or at the time of an event (e.g. when a device is plugged into the OBD port, upon configuration changes, when Universal Diagnostic Services or J2534 initiates a memory-modification event or software reflash, at a set mileage or time interval, upon receipt of a [digital] service notice, etc.) using the in-vehicle eSIM and telemetry module, or during vehicle servicing when the programming device is directly connected to a wide-area network. The dual-use of in-car and service devices to update the master blockchain eliminates the challenge of home-service events going unrecorded. Each vehicle OEM could therefore track all of the light clients' data that is reported to their blockchain node so that they can monitor their vehicles for potential risks. This addresses software validation during both over-the-air and service-center reprogramming events. The OEM would then be able to provide servicing companies insights about necessary updates for given vehicles in a fleet.

\subsection{Level 1}
All networked in-vehicle embedded devices will be responsible for generating a hash of the software version, date and time at pre-determined intervals or events. The hash need not be limited to this use case and its associated data. The status of other security-relevant software components could also be collected such as the status of V2V authentication certificates. This will be stored in a node of the DHT which would reside on one of the embedded devices in the vehicle. The DHT node will connect to other embedded systems with their own DHT node within the car's internal network.

We used DHT as a means for data storage because of the distributed, redundant and scalable features of the ledger. A DHT does not have the same financial cost requirements as using a public blockchain, nor does it require public Internet access to use. DHT requires considerably less memory overhead than a blockchain node, where in some cases as little as 1-2KB of ROM is needed at each node as compared with over 1MB for even the smallest blockchain light clients \cite{the}.

Further, a user may retrieve the hash of any embedded device on the vehicle given they have the public key to the hash table. The hash key would be computed from some irreversible combination of vehicle parameters to complicate deanonymization. One potential key-generation approach might take as input a feature vector describing the serial numbers of each ECU in the vehicle, and the most-recent software version within the vehicle so that the key captures critical vehicle parameters, and still changes over time (to further mitigate the effects of key deanonymization). The user may therefore be able to determine their key through an associated mobile application provided by the OEM, or else generate it through a web interface by inputting their VIN and a unique passphrase provided by the OEM once the user's ownership has been validated, similar to the process for retrieving an immobilizer or key code. This approach reduces the potential for third-party access to sensitive data, and may serve to limit abuse. 

The benefit of this key-based approach is that if a device is being unresponsive, the last known hash of the embedded device can be checked from other DHT participants. This is a benefit of DHT's decentralized and fault-tolerant nature. An additional benefit is that this approach can help alert an operator that a single module has been modified without consistency across the vehicle, e.g. that the odometer's mileage was changed in the engine controller but not the transmission controller. In such cases, discrepancies can be flagged using pre-defined logic or manually, through human-assisted inspection. Considering there may be hundreds of embedded devices per vehicle, a DHT's ability to scale efficiently is incredibly valuable in assuring system integrity in such complicated and hyper-connected systems.

\subsection{Level 2}
The local DHT will also be housed on a master unit, such as the infotainment head unit.  The master unit will take a hash of the DHT - a meta hash of all software versions across the embedded systems on the vehicle - and hash this onto a blockchain light client in regular intervals (while also storing a copy on the master unit). Upon capturing this meta hash, the DHT nodes will be directed to erase their oldest stored values once a storage limit is reached, in order to preserve memory in the embedded devices. The number of values to store must be selected carefully to balance the memory consumption, hardware longevity, and data integrity of the system (e.g. storing more values provides resilience in the case of data corruption within the master node, but requires more [costly] onboard memory, while reducing storage saves cost, but frequent rewrites may compromise memory integrity as the hardware write-limit is neared.  

We note that in cases, it may be desirable to maintain a degree of redundancy within the system's data storage at the DHT level, for example in the case where connectivity is intermittent and data may not be uploaded to the blockchain at regular intervals. In such a scenario, corruption of one DHT node could result in loss of information to the blockchain at the next upload event. Redundancy within DHT has the potential to avoid this data loss, which may be of critical importance in high-assurance software implementations.

One possible solution for avoiding DHT data loss or corruption relies on the use of clusters of redundant DHT nodes within the vehicle. Rather than storing the full DHT from information derived from a single client, multiple clients may be considered in aggregate to ensure data integrity. The number of clients may be determined as a function of memory, cost longevity, and resilience targets.

\begin{figure}[h!]
\centering
\includegraphics[width=3in]{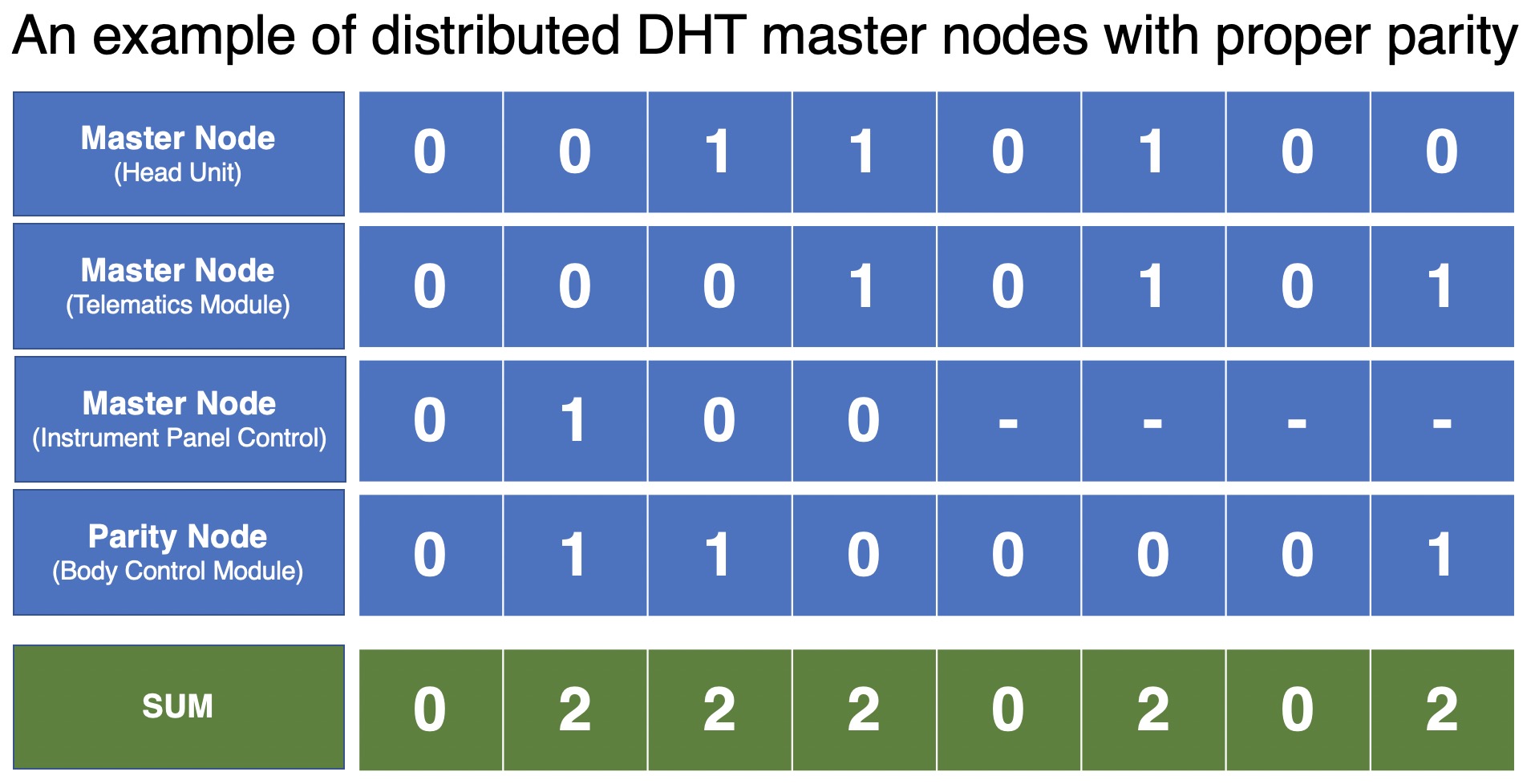}
\caption{A parity system stores data across multiple storage devices. Bits are added across devices and ``checked'' against a parity device. If there is a disparity between expected and actual values, any single device's memory can be reconstructed. Storage size may differ across devices, but the parity storage must be the largest storage device in the system.}
\label{fig:parity}
\end{figure}

One representative implementation to maximize data resilience is to implement at least two DHT nodes with a third node having equivalent data storage serving as a parity disk. A parity process is designed such that a single-bit change across a set of bits sums to a known value, e.g. with the sum of values across a given bit-location for all input disks summing to an even number. With such a system, rather than having every compute module capable of supporting a DHT node operating independently, they would instead form meta-client clusters of three or more units with heightened resilience. That is, two of the three modules may be in agreement and therefore reconstruct lost or corrupted data. In this manner, disparities in values across devices may be detected and corrected, providing protection against data corruption or modification and reinforcing the layered and distributed nature of the presented black box architecture. A representative implementation of such a distributed parity system appears in Figure~\ref{fig:parity}.

We assume that the master unit will have more memory and processing power than typical embedded systems in the vehicle and can accommodate such a light client, which is reasonable based on the architecture of contemporary vehicles, which have one or more high-power, high-compute nodes bridging across intra- and extra-vehicular networks. These devices may have several megabytes or even gigabytes of RAM and ROM, as well as multi-core processors. 

\subsection{Level 3}
The head unit will feed the light client's meta-hash to the full blockchain node designated for the light client. The full node will be run by the OEM of the vehicle, or another fleet management or agency. Once the light client delivers the meta-hash to the full node, a checksum will be conducted on the OEM's server. The OEM will have a library of meta-hashes that reflect the acceptable and safe permutations of embedded software versions across all possible vehicle variants. If the vehicle's meta-hash fails the checksum, a notification will be sent to the vehicle service provider indicating service is needed. Alternatively, an emergency over-the-air update could be delivered for all embedded software in the vehicle, or the vehicle could be immobilized until the software has been manually reinstalled. 

A benefit of employing the blockchain once data are communicated over the open Internet is the integrity afforded by a large, public blockchain like Ethereum, which would only be vulnerable to a $51$\% attack. A $51$\% attack can only occur if an attacker manipulated over $50$\% of the compute power of the blockchain and diverts the chain's direction by altering the values of the blocks for the majority of nodes. For large blockchains such as Ethereum or Bitcoin, this is extremely expensive and difficult to accomplish considering the amount of compute power behind the chain. 

\begin{figure}[h!]
\centering
\includegraphics[width=4in]{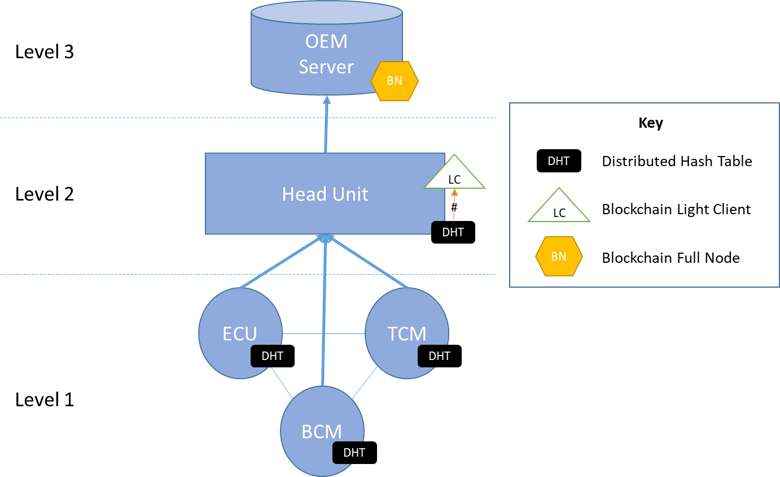}
\caption{This three-layered architecture blends a DHT implementation with a blockchain implementation, to gain the benefits of both technologies without imposing undue computational burden on constrained compute modules. It shows a representative but simplified system featuring an Engine Control Unit (ECU), Body Control Module (BCM), Transmission Control Module (TCM), and Head Unit. The Head Unit acts as a master node wherein the DHT results are meta-hashed and hashed onto a public blockchain.}
\label{fig:node}
\end{figure}

While we acknowledge that our solution is best implemented as a clean-sheet design, we targeted the capabilities of contemporary vehicle architectures to ensure our implementation's feasibility on low-cost, constrained vehicle hardware. Were such a system to be implemented on an already-deployed vehicle, first, one must ensure that the hardware payload offered appropriate capabilities related to computation, data storage, and secure communications. Then, existing software must be redeveloped with the proposed solution in mind. In redeveloping software, safety and security implications must be considered and reevaluated (e.g. ISO26262\cite{iso26262}), and in many cases, the cost of implementation may not be deemed worthwhile.

\section{Alternative Approaches}
Previously mentioned design requirements include 1) self-identification of software and hardware metadata hash identifiers to a master computing node, 2) communication integrity of the embedded system metadata and 3) the master unit holding version control must be able to integrally communicate with other high-compute units over wireless telemetry. The proposed DHT/Blockchain hybrid successfully meets these requirements in the following ways: 1) the DHT nodes for each embedded system can be called by the light client stored in the master computing node for each vehicle; 2) the public blockchain's resilience to attack and manipulation through its decentralized, consensus-driven cryptographic hashing function provides integrity of the vehicle metadata that represents the aggregate of the embedded system software details; and 3) the proven security behind public blockchains like Ethereum and its ability to withstand attacks is capable of providing integral communication with other high-compute units. Alternative approaches such as using exclusively DHT or blockchain would not address all three requirements. 

A DHT-only solution would violate requirements 2 and 3. DHTs are prone to a wide variety of attacks that could compromise the integrity of the data within \cite{urdaneta2011survey}. Even if additional integrity measures were applied, such as employing a Merkle-tree for hash verification, routing attacks may still be feasible. Instead, the security and tamper-proof benefits of the blockchain are valuable once the data extends outside of the vehicle's intranet. Also, DHT would complicate requirement 3. With DHT, data is distributed amongst associated peer nodes - no node holds the complete ledger - and therefore, because each DHT node only contains a portion of the routing table, the OEM would need to contact multiple nodes in order to find the locations and values for the entire database. Considering the scale at which these auditing operations would need to occur, coalescing all necessary data on the blockchain simplifies the OEM's data retrieval for their fleet. Each master node will hold the complete DHT database so that the OEM could access all relevant vehicle data in the same location through a single address - rather than having to query all embedded devices individually. Therefore, while design requirement 3 is feasible to accomplish with DHT, it is highly inefficient. 

Initially, employing a blockchain light client was envisioned for each embedded device's reporting log. The benefits of using a blockchain would be manifold; blockchains have integral communications, are fully distributed, and offer data immutability. However, blockchain alone would violate requirement 1 of the design considerations. Because of the sheer memory space and compute power required, it would not be feasible for the edge embedded systems to host a blockchain node or even a light client that could deliver a hash identifier to another node. Even light client blockchain nodes are around 1MB in size and it is infeasible to believe that all contemporary and legacy in-vehicle embedded devices could host a node of such a size\cite{neuromesh}. Further, the target embedded systems have a constrained clock cycle, limited extra-vehicular bandwidth and RAM limitations. Further, many systems are optimized for hard-realtime applications and maximally utilize available resources, which would make employing blockchain for all nodes difficult.

To receive the benefit of immutability, an existing major blockchain such as Bitcoin or Ethereum would need to be used so that it is sufficiently large that an attacker would have trouble achieving a $51$\% attack, which is why a private blockchain that does not require a connection to the internet would be insufficient. Further complicating a blockchain-only approach, each embedded node would need to directly connect to the Internet to hash values to a public blockchain (while private blockchains may be used, and indeed preferred by some OEMs, the use of public blockchains would enhance trust and potentially enable novel applications and data uses, e.g. by insurers, used vehicle certification services, or government agencies for VMT applications\cite{SiegelMasters} or law enforcement). This would come with additional compute and cost overhead that could make such an architecture operationally infeasible.

Together, DHT and Blockchain are able to meet all three design considerations. A hybrid approach of DHT and blockchain has merit for implementing a secure, scalable, and computationally-efficient record of software versions. Additional benefits of using the blockchain is that while the in-vehicle DHT may store the current software payload, the blockchain may provide a historical record of changes. This has the potential to be useful for insurers in accident reconstruction, Information Sharing and Analysis Center (ISAC) industry members that could benefit from understanding how an attack on vehicle data or software unfolded\cite{falco2019cyber}, or OEMs, in assessing warranty claims, among other applications. 

It is worth noting that with regular data uploads from a vehicle to an OEM's Cloud, privacy rightly becomes a concern. While the intended design of the DHT/blockchain system is to report data related to vehicle configuration and operation, rather than directly related to user habits, and to do so in an anonymous manner, it is na\"{i}ve to assume that such protections will be enduring and sufficient to assure user privacy. Therefore, OEM privacy policies must be modified to reflect this new source of potential data leakage. Signing a policy is common upon purchasing a vehicle, and vehicle occupants may agree to certain policies implicitly, simply through their engaging with the vehicle. While the technical solution we propose aims to preserve user privacy and overall enhance vehicle security, ultimately, it is up to the OEMs to ensure that they have informed consent from their stakeholders\cite{tos_paper}. 

\section{Enabled Capabilities}\label{results}
The DHT and blockchain-light implementation will enable a safer and more adaptable vision for future vehicle software assurance. As described, we primarily address issues of software version control, but also may store critical data such as vehicle identification or odometer data in a similar manner, updating the DHT and blockchain upon key event types. Eventually, this architecture may also be used to assure the integrity of critical sensor or actuator data and commands, once in-computer hardware is sufficiently capable. 

This distributed black box approach allows OEM's and fleet managers to ensure that a vehicle is running the latest version of software from an approved vendor, that the hardware has been maintained appropriately, and that there have been no unauthorized modifications that could lead to inconsistencies or unanticipated/untested behavior. It also ensures that vehicles are not reprogrammed to extend warranties, in fraudulent attempts to increase resale value, or to misrepresent vehicle options. 

For deployed vehicles, it increases the likelihood that OEMs and suppliers can identify vehicles subject to software or hardware recalls, and for insurers, to ensure that the software within the vehicle is authentic and has not been modified by a user, leading to its potential failure. 

Additionally, the DHT may be used as one element in a larger system for ensuring the existence of a secure computing environment within the vehicle, by comparing the stored blocks to ensure that all of the components that are supposed to be present are there and have not been altered, and that historic events have not been tampered with. Validating data integrity, even weakly, reduces the likelihood that modules can be added, removed, or replaced without proper authorization (a service event where the chains are told to expect new hardware on the next boot -- protected by a secure, rolling and encrypted seed/key unlock to prevent EEPROM dump attacks), and helps to assure that critical data -- such as the VIN, odometer, or crash history -- have not been altered (since it would require compromising the majority of the chain at the same time). Disagreement could indicate probable tampering or fraud and set a flag across all modules that remains until being cleared by a proprietary tool. While compromising such a system is possible, it would take heightened effort relative to today's security implementations, and may serve as a deterrent from less-determined adversaries. 

Integrating the ledgers as a low-level security protocol (validating the presence of other modules and the contents of their chains at startup) may prevent incidents stemming from vehicle communication spoofing (e.g. taking a DSRC module from a scrapped car and moving it to an adversarial vehicle or standalone device to simulate congestion and unnecessarily cause traffic). 

With a public ledger, it will be possible to identify when a vehicle's mileage is reset, or when the airbag light is turned off, or that a car did (not) have its recalled software updated or otherwise reconfigured. We can also see that modules have been replaced or reprogrammed, e.g. with one from a junkyard, enabling a smarter ``CarFax'' like service backed by authentic data from real vehicles. This is of particular value to individuals and organizations concerned with temporal accountability, such as insurers, manufacturers, suppliers, lenders, and law enforcement agencies. As vehicle automation increases and manufacturer accountability grows, this approach will further support the creation of auditable records for vehicle use. 

Implementing the distributed black box at scale will require OEM buy-in to update in-vehicle software to support DHT and blockchain-light, particularly in the incumbent fleet. It will also require the deployment and management of improved version control databases and enhanced testing of software and hardware compatibility. While readers may question why OEM's and suppliers would go through the time and cost hassle of implementing such a system, the reality is that software is increasingly a competitive advantage differentiating vehicles within a segment. Implementing a DHT-blockchain hybrid within vehicles will allow OEM's to safely update in-vehicle software and ensure that vehicles are maintained appropriately over time, easing both the rollout of comfort and convenience features, but also the long-term deployment of high-automation and self-driving capabilities, without increasing liability exposure.

While some OEMs already update software over the air, and while some OEMs implement solutions such as Uptane and other PKI-based solutions have been proposed\cite{kent_paper} to ensure that firmware payloads are secure, it is unclear whether these OEMs take adequate measures to ensure that the appropriate software versions are installed to a vehicle and capable of interoperating with one another. The addition of a blockchain solution, updated upon hardware change events, has the potential to remove ambiguity and ensure that the right software is installed with respect to the vehicle's hardware payload and aggregate set of software variants. Without fully-updated blockchain, traditional update approaches may fail - e.g. in the case that a hardware module was replaced by a repair shop or end-user without notifying the OEM of the change. Such an update could result in bricking the device, or, arguably worse - the vehicle may operate normally until some edge case not tested for that particular variant is encountered. The ability to \textit{safely} update software and ensure system-level compatibility will also enable long-term annuity models for revenue capable of augmenting time-of-sale and traditional service models. 

The distributed black box and other technologies will be essential in helping transition towards highly-automated vehicles and growing the role of shared mobility services. 

\section{Future Opportunities}\label{opportunities}
Vehicle computation will only get faster and networks will grow more complicated with the advent of highly-automated and increasingly connected fleets\cite{siegel_itsj}. Enabling this vision, modules will increasingly gain compute capability including more and faster cores. Networks -- both in-vehicle and extra-vehicular -- will gain capacity and latency will drop, ROM and RAM will increase, and software complexity will continue to grow as critical functions are further relegated to computer-control and additional comfort-and-convenience features are added at time of sale and in aftersale operation. These advances will be especially prevalent in shared mobility vehicles, for which the initial costs may be higher as they are not borne as directly by occupants. 

As highly-performant vehicles gain marketshare and older vehicles atrophy and are removed from the operational fleet, we can advance the capabilities of the vehicular ledgers to support more resilient implementations and broader applications for authenticated in-car data, including generating data useful for high-criticality network-wide fleet optimization.  

The automotive software provenance solution as described may be extended to map not only vehicle software version, configuration, and changes, but other parameters as well -- for example to validate that the tires were changed at a particular time by a particular vendor, or that the air filter was replaced when required by the on-board diagnostic system. Ultimately, there may exist a blockchain dedicated to monitoring wholesale vehicle maintenance and operations.

Such use cases lend themselves to a non-repudiable way of reducing risk exposure across all elements of the product life cycle, thereby empowering insurance with reliable data, reducing warranty costs by minimizing false claims and authenticating systems at the time of manufacturing, use and service.

Future implementations should be tested for specific use cases. For example, DHT/blockchain at each wheel encoder could ensure that odometer increments as it should based on vehicle motion patterns, enabling more accurate Vehicle Miles Travelled (VMT) tax or Usage Based Insurance (UBI)\cite{SiegelMasters}. This could feed into the concept of ``Proof of Location'' for connected vehicles, wherein vehicles generate a reliable history of location over time to prove that they were (not) present in a particular location, and to bear witness to (and provide data about) traffic incidents, road conditions and weather events.

As the automotive industry continues moving towards high automation and full self-driving, there will be increasing scrutiny of the reliability of the embedded systems, their data and software. Enabling these cyber-physical systems with a distributed black box audit trail that provides operational transparency and integrity will help to assure users and OEMs that systems are operating as intended. 

Realizing the full potential of highly-automated systems and their associated adoption requires human trust in their operations. As part of building trust, such systems will need to be monitored and maintained for safety and security. Just as there were risks associated with civil aviation when first introduced to the public, the closely managed governance processes, which included the introduction of Flight Data Recorders, facilitated their wide acceptance. The highly-automated vehicle is at a similar critical juncture to the introduction of civil aviation where processes, procedures and technology will be needed to assure the public of their reliability. A black box audit trail is one such technology. The merit of a black box audit trail has been discussed at length over the past five years, but the inherent connectivity and associated security concerns of highly-automated systems necessitates that the design differs from traditional FDRs. The proposed distributed black box system is a layered, resilient, and robust approach to assuring the integrity of connected and automated vehicles and other autonomous systems. Implemented properly, it has the potential to enable high-assurance systems and improve auditability for OEMs, suppliers, vehicle operators, and other stakeholders and drive the safe, large-scale adoption of next-generation automotive technologies.

\bibliographystyle{unsrt}  
\bibliography{./bibliography/IEEEabrv,./bibliography/IEEEexample}
\end{document}